\documentstyle[12pt]{article}

\def\appendix#1{
  \addtocounter{section}{1}
  \setcounter{equation}{0}
  \renewcommand{\thesection}{\Alph{section}}
  \section*{Appendix \thesection\protect\indent \parbox[t]{11.715cm}
{#1} }
  \addcontentsline{toc}{section}{Appendix \thesection\ \ \ #1}
  }

\let\a=\alpha\let\b=\beta\let\d=\delta
\let\e=\epsilon\let\g=\gamma

\let\s=\sigma

\let\G=\Gamma

\let\vt=\vartheta
\newcommand{\nn}{\nonumber}
\newcommand{\be}{\begin{equation}}
\newcommand{\ee}{\end{equation}}
\newcommand{\bea}{\begin{eqnarray}}
\newcommand{\eea}{\end{eqnarray}}

\newcommand{\tP}{\tilde{P}}
\newcommand{\tQ}{\tilde{Q}}
\newcommand{\tS}{\tilde{S}}
\newcommand{\tD}{\tilde{D}}
\newcommand{\tK}{\tilde{K}}

\newcommand{\nbox}{{\,\lower0.9pt\vbox{\hrule \hbox{\vrule height 0.2 cm
\hskip
0.2 cm \vrule height 0.2 cm}\hrule}\,}}

\begin{document}
\thispagestyle{empty}

\begin{flushright}
\hfill{SU-ITP-98/55}\\
\hfill{RU-98-41}\\
\hfill{hep-th/9809164}\\
\hfill{September 1998}\\
\end{flushright}

\vspace{20pt}

\begin{center}
{\large {\bf The GS String Action on $AdS_3\times S^3$
with Ramond-Ramond Charge}}

\vspace{40pt}

{\bf J. Rahmfeld$^{a}$ and Arvind Rajaraman$^{b}$}

\vspace{20pt}
$\phantom{a}^a${\it Department of Physics,

Stanford University,

Stanford, CA 94305-4060}

\vspace{20pt}
$\phantom{a}^b${\it Serin Physics Lab,

Rutgers University,

Piscataway NJ, 08854}

\vspace{60pt}

\underline{ABSTRACT}

\end{center}

We derive the classical $\kappa$-symmetric Type IIB string action
on $AdS_3\times S^3$ by employing the $SU(1,1|2)^2$ algebra.
We then gauge fix $\kappa$-symmetry in the background adapted
Killing spinor gauge and present the action in a very simple
form.

{\vfill\leftline{}\vfill
\vskip  30pt
\footnoterule
\noindent
{\footnotesize
$\phantom{a}^a$ e-mail: rahmfeld@leland.stanford.edu}  \vskip  -5pt

\noindent
{\footnotesize
$\phantom{b}^b$ e-mail: arvindra@alumni.stanford.org}  \vskip  -5pt


\pagebreak
\setcounter{page}{1}

\section{Introduction}

There has been great interest recently in string theory
on $AdS_5 \times S_5$ \cite{Malda, Klebanov, Witten}
due to its possible relation to ${\cal N}=4,d=4$
Yang-Mills theory. Whereas the large $g^2N$ limit is conjectured to
be dual to Type $IIB$ supergravity on this manifold,
for which there is by now mounting evidence, stringy effects
are supposed to correspond to $1/g^2N$  corrections \cite{Malda} in the
Yang-Mills theory.
It is of great interest, therefore to construct string theory in this
background. Although there has been significant progress in this direction
\cite{Tseytlin, NearHorizon,SuperKilling,
ads5, realSimple}, the action (so far)
has proven too difficult to quantize.
In this note, we will try to analyze a simpler case, that of
string theory on $AdS_3\times S^3$.


One interesting aspect of this background
is that the compactification of $D=6$ supergravity on $S^3$ can
be achieved in two fundamentally different ways: the charged
three-form field strength can either be of NS or RR type.
In \cite{Giveon} the NS field was charged and a significant
understanding of a string propagating in this background was
achieved. In this paper we focus on a string in the non-trivial
RR background and construct the string action in the Green-Schwarz
(GS) formulation \cite{GreenSchwarz}. The hope
is that eventually this case can be better understood, maybe by
relating it to results of \cite{Giveon}.
Various other aspects of this background have been studied in
\cite{juan,klaus,ergin,deBoer,mike}.

We shall follow the approach of \cite{Tseytlin} which requires
a description of the background as a supercoset manifold.
The $AdS_3\times S^3$ background is the near-horizon geometry
of the $D1-D5$ brane system and is a solution of chiral
$N=2 \ (2,0)$ supergravity in six dimensions \cite{Romans}
preserving all 16 supersymmetries. By essentially straightforward
extension of the arguments given in \cite{exact} it can be shown
that the solution does not get any $\a'$ corrections which is
a necessity to formulate string theory in this background.
In \cite{conformal} it was noted that the isometry group of $D1-D5$ system
is $SU(1,1|2)^2$, and hence the background can be viewed as the supercoset
space $\frac{SU(1,1|2)^2}{SO(1,2)\times SO(3)}$.
The construction of the action following \cite{Tseytlin}is then
straightforward except for the construction of the
Wess-Zumino term, which
requires some trial and error.


This is done in section two, where we
start with  the algebra of $SU(1,1|2)^2$ (which we
derive very explicitly in 6D covariant form from the
$SU(1,1|2)$ algebra in appendix A) and construct
the Wess-Zumino term from first principles following
\cite{Tseytlin}. We find, in fact a continuous family
of WZ terms interpolating between the pure NS background
and the RR background.

The resulting
GS action is then given in terms of supervielbeins
which we also solve for in section three. In section four
we gauge
fix $\kappa$-symmetry in the `background adapted Killing
spinor gauge' \cite{FixKilling,SuperKilling, ads5} which simplifies
the action considerably.

Finally we present our conclusions and some open questions.

\section{From the Algebra $SU(1,1|2)^2$ to the String Action on
$AdS_3\times S^3$}
\setcounter{equation}{0}

The target space of
string theory on $AdS_3\times S^3$ with 16 supersymmetry generators
is the supercoset manifold
$\frac{SU(1,1|2)^2}{SO(1,2) \times SO(3)}$ whose bosonic part
is $\frac{SO(2,2)\times SO(4)}{SO(1,2) \times SO(3)}$. The
generators of this supergroup are the momenta and Lorentz
transformations on $AdS_3$ and $S^3$
\be
P_{a},J_{ab}, \ {\rm and } \ P_{a'},J_{a'b'}
\ee
where $a=0,1,2$ and $a'=3,4,5$, plus 2 complex chiral
$6D$ spinors
\be
Q_{I\a\a'}\otimes \pmatrix{1\cr 0} \label{6dcharge}
\ee
with $I=1,2,\ \a=1,2,\ \a'=1,2$.  Our conventions are
\be
\G^a=\g^a\otimes 1 \otimes \s_1, \qquad \G^{a'}=1\otimes
\g^{a'}\otimes \s_2 \label{gammamat}
\ee
where $\g^0=i\s^3, \g^{1,2}=\s^{1,2}, \g^{a'}=\s^{a'-2}$.
In the following we will freely use $\g^a$ short for  $\g^a\otimes
1$ (and the same for primed indices).
With these definitions it is clear that $Q_I$ defined as
above is indeed chiral. The conjugate supercharge $\bar Q^{I\a\a'}$
is defined by
\be
\bar Q^{I}=(Q^I)^\dagger \g^0.\label{barQ}
\ee
Crucial for the construction of the action are
the (antihermitean) supervielbeins
$L^a,L^{a'}, L^I$ and $\bar L^I$ and the superconnection
$L^{ab}$ and $L^{a'b'}$. Being a $\s$-model with a supercoset
as the target space, the action is only allowed to contain
the supervielbeins. and will be of the  general structure
\be
S=S_{kin}+S_{WZ}.
\ee
The kinetic term is next to trivial to write down, the
more subtle issue is to construct the Wess-Zumino term,
needed for $\kappa$ invariance,
which is an integral over a closed 3-form. To find this
form we need the superalgebra, derive from there the
Maurer-Cartan equations and identify a unique closed
three-form built from the supervielbeins.

It should be apparent by now that an important ingredient
is the $SU(1,1|2)^2$ algebra
$$
\{Q_I, \bar Q_J\}=2\d_{IJ}\left(iP_{a}\gamma^a -
P_{a'}\gamma^{a'}\right) +\e_{IJ}\left(J_{ab} \gamma^{ab} -
J_{a'b'} \gamma^{a'b'}\right) \nn
$$
\vspace{-8mm}
\bea
\left[P_{a},Q_I\right]=-\frac{i}{2} \e_{IJ}\gamma_a Q_J  & &
\left[P_{a'},Q_I\right]=\frac{1}{2} \e_{IJ}\gamma_{a'} Q_J\nn \\
\left[J_{ab},Q_I\right]=-\frac{1}{2} \gamma_{ab} Q_I &&
\left[J_{a'b'},Q_I\right]=-\frac{1}{2} \gamma_{a'b'} Q_I  \nn \\
\left[P_{a},\bar Q_I\right]=\frac{i}{2} \bar Q_J \e_{JI}\gamma_a
&& \left[P_{a'},\bar Q_I\right]=-\frac{1}{2} \bar Q_J \e_{JI}
\gamma_{a'} \label{algebra}\\
 \left[J_{ab},\bar Q_I\right]=\frac{1}{2}
\bar Q_I\gamma_{ab} && \left[J_{a'b'},\bar Q_I\right]=\frac{1}{2}
\bar Q_I \gamma_{a'b'} \nn \eea
\vspace{-8mm}
\bea
\left[M_{AB},M_{CD}\right]&=&\eta_{BC} M_{AD}+\eta_{AD}
M_{BC}-\eta_{AC} M_{BD}-\eta_{BD} M_{AC}\nn \\
\left[M_{A'B'},M_{C'D'}\right]&=&\d_{B'C'} M_{A'D'}+\d_{A'D'}
M_{B'C'}- \nn \\ & & \hspace*{2cm} -\d_{A'C'} M_{B'D'}-\d_{B'D'}
M_{AC} \nn
\eea
where we defined
\be
P_{a}=M_{0a}, \ J_{ab}=M_{ab}, \ P_{a'}=M_{0'a'}, \
J_{a'b'}=M_{a'b'} \label{PJM}
\ee
and where $\eta=(-++-)$. Note that the bosonic generators
are taken to be anti-hermitean.
The defining equations of the background can be obtained
in the standard way by defining the group derivative
\be
{\cal D}=d + L^a P_a +\frac{1}{2}L^{ab} J_{ab}+L^{a'} P_{a'} +
 \frac{1}{2}L^{a'b'} J_{a'b'}+\frac{1}{2}(\bar Q^I L_I +\bar L^I
 Q_I)
 \ee
and requiring ${\cal D}^2=0$. This leads to
the Maurer-Cartan equations:
\bea
dL^a&=&-\frac{i}{2}\bar L^I\g^a\wedge L^I -L^b\wedge L^{ba} \nn \\
dL^{a'}&=&+\frac{1}{2}\bar L^I\g^{a'}\wedge L^I -L^{b'}\wedge L^{b'a'}
\nn \\
dL^I&=&-\frac{i}{2}\e^{IJ}L^{a}\wedge\g_{a} L^J +
        \frac{1}{2}\e^{IJ}L^{a'}\wedge\g_{a'} L^J - \nn \\
        & & \qquad \qquad
-\frac{1}{4}L^{ab}\wedge\g_{ab} L^I -
\frac{1}{4}L^{a'b'}\wedge\g_{a'b'} L^I \label{Maurer} \\ d\bar
L^I&=&\frac{i}{2}\e^{IJ}\bar L^{J}\g_{a}\wedge L^{a} -
\frac{1}{2}\e^{IJ}\bar L^{J}\g_{a'}\wedge L^{a'}- \nn \\
        & & \qquad \qquad
-\frac{1}{4}\bar L^I \g_{ab}\wedge L^{ab}
-\frac{1}{4}\bar L^I \g_{a'b'}\wedge L^{a'b'} \nn
\eea
plus the non-relevant ones for $dL^{ab}$ and $dL^{a'b'}$.
The Wess-Zumino term can be constructed in terms of the
vielbeins without actually solving these equations.  In the
background at hand this
is only slightly more subtle than in the $AdS_5\times
S^5$ background, since the $L^I$ do not obey any Majorana conditions.
We find that the unique form satisfying the requirements is
\bea
{\cal H}_3&=&As^{IJ}\left(\bar L^I \g_a\wedge L^J\wedge L^a+i\bar L^I
\g_{a'} \wedge L^J\wedge L^{a'}\right)+c.c. \nn\\
&=&As^{IJ}\left(\bar L^I \G_a\wedge L^J\wedge L^a+\bar L^I
\G_{a'} \wedge L^J\wedge L^{a'}\right)+c.c.
\eea
where $s^{IJ}=\s_3^{IJ}$.
In proving that $d{\cal H}_3=0$ one has to apply the identities
(\ref{Fierz}), and has to use
\be
s^{IJ}\left(\bar L ^I \g_a L^J\bar L ^K \g^a L^K-
\bar L ^I \g_{a'} L^J\bar L ^K \g^{a'} L^K\right)=0.
\ee
It remains to find the coefficient in front of the Wess-Zumino
term. For this we consider the flat-space limit, where
the vielbeins read in our notation (see \ref{vielbein} with
${\cal M}=0$ and $s=1$):
\bea
  L^I&=&d\theta^I \nn \\
  \bar L^I&=&d\bar \theta^I \nn \\
  L^a&=&dx^a-\frac{i}{4}
         \left(\bar\theta^I \g^a d \theta^I-d\bar\theta^I \g^a
         \theta^I\right)\\
  L^{a'}&=&dx^{a'}+\frac{1}{4}
         \left(\bar\theta^I \g^{a'} d \theta^I-d\bar\theta^I
         \g^{a'}
         \theta^I\right) \nn
\eea
Therefore,
\bea
WZ&=& A s^{IJ}\bar L^I\G_a\wedge L^J\wedge L^a +c.c \nn \\&=&A \
(d\bar\theta^1 \G_a d\theta^1-
d\bar\theta^2 \G_a \wedge d\theta^2)
\wedge dx^a+ c.c +...
 \nn \\
&=&A\ d\Biggl(\biggl(
(\bar\theta^1 \G_a d\theta^1-d\bar\theta^1 \G_a \theta^1) - \nn \\
&& \quad \qquad-(\bar\theta^2 \G_a \wedge d\theta^2-d\bar\theta^2 \G_a
\wedge
\theta^2)\biggr)
\wedge dx^a+...\Biggr) 
\partial_\b\theta^1-\partial_\a\hat\theta^2
\eea
and hence
\be
\int_{M^3} WZ=A\int_{M^2}d^2\s \ \e^{ij}(\bar\theta^1 \G_a
\partial_i\theta^1- \partial_i\bar\theta^1 \G_a
\theta^1)\partial_j x^a+.... \ee
By comparison with standard literature (see for example \cite{GSW})
one finds
\be
A=\frac{i}{4}.
\ee
Therefore, the $6D$ superstring action is given by
\bea
S&=&-\frac{1}{2}\int_{M^2}d^2\s \ \left(L^a L^a+L^{a'}L^{a'}\right)+ \nn
\\
&&  +
\frac{i}{4}\int_{M^3}s^{IJ}\left(\left(\bar L^I \g_a\wedge L^J\wedge
L^a+i\bar L^I
\g_{a'} \wedge L^J\wedge L^{a'}\right)+c.c.\right) \label{action}\\
&=&-\frac{1}{2}\int_{M^2}d^2\s \ \left(L^a L^a+L^{a'}L^{a'}\right)+ \nn
\\
&&  +
\frac{i}{4}\int_{M^3}s^{IJ}\left(\left(\bar L^I \G_a\wedge L^J\wedge
L^a+\bar L^I
\G_{a'} \wedge L^J\wedge L^{a'}\right)+c.c.\right), \nn
\eea
which is the main result of this section.

This WZ term, however, should not really be unique, since there
exists also the string in the same geometry, but charged under
the $NS$ $B$-field, and there must be a different WZ term for it.
The answer suggested by the work of \cite{BST} answer is that the
general WZ term should be given by
\be
{\cal H}\sim s^{IJ}\left(\left(\bar L^I \g_a\wedge L^J\wedge L^a+i\bar
L^I
\g_{a'} \wedge L^J\wedge L^{a'}\right)+c.c.\right)+L^a\wedge L^b
\wedge L^c H^+_{abc},
\ee
where $H^+_{abc}$ is one of the five components of the self-dual
superfield
\cite{awada}. This is to be understood from the point of
view of compactifying the $D=10, N=2$ Type IIB theory on $K3$ (and
truncating the matter fields). Of the five self-dual field
strengths that arise \cite{Triality}, three find their origin in the
self-dual five-form field strength, one from the RR three-form (plus its dual)
and one ($H^+_{abc}$) from the NS three-form and its dual in $D=10$.

\section{The Supergeometry}
\setcounter{equation}{0}

It remains to actually solve the Maurer-Cartan equations and
obtain the supervielbeins. The general method is standard
and was outlined for example in \cite{Tseytlin} where for the
$AdS_5\times S^5$ case the vielbeins were constructed up to
quartic order. In \cite{NearHorizon} it was observed that the
equations can in fact be integrated and the supergeometry can
be found in closed form.

To do so we  have to play the usual trick and introduce
$\theta_s=s\theta$ to solve for a generalized
vielbein $L_s$ from which one obtains eventually the standard
vielbein as $L=L_{s=1}$. In the process  we also find following
\cite{D3} a convenient form of the Wess-Zumino as
a two-dimensional worldsheet integral, integrated once more
over the parameter $s$.

Let us denote the general structure of the algebra by
\bea \{Q_I, \bar Q_{\bar J}\} & = & f_{I\bar J}^A B_A \nn \\
\left[B_A,Q_I\right] & = & f_{AI}^J Q_J \nn \\
\left[B_A,\bar Q_{\bar I}\right] & = & f_{A\bar I}^{\bar J}
\bar Q_{\bar J}  \\
\left[B_A,B_B\right] & = & f_{AB}^C B_C,\nn
\eea
where the distinction between $I$ and $\bar I$ serves only
the purpose to keep track of $Q$ and $\bar Q$.
With $D$ being the standard covariant (bosonic) derivative
\be
D=d+\frac{1}{4}\omega^{ab}J_{ab}+\frac{1}{4}\omega^{a'b'}J_{a'b'}
 +e^aP_a + e^{a'} P_{a'}  \label{D}
\ee
we find from
\be
e^{-\frac{s}{2}\left(\bar \theta Q+\bar Q \theta \right)}D
e^{\frac{s}{2}\left(\bar \theta Q+\bar Q \theta \right)}=
{L}_s^A B_A+\frac{1}{2}\left(\bar {L}_s Q+\bar Q L_s \right)
\label{prop}
\ee
the differential equations
\bea
\partial_s L^A_{s} & = & -\frac{1}{4} \bar \theta^I f_{I\bar J}^A
L_s^{\bar J}+\frac{1}{4} \bar L_s^I f_{I\bar J}^A \theta^{\bar J} \nn
\\
\partial_s L^{\bar I}_{s} & = & d\theta^{\bar I} +L_s^B f_{B\bar
J}^{\bar
I}\theta^{\bar J} \label{diffs} \\
\partial_s \bar L^I_{s} & = & d\bar \theta^{I} -\bar \theta^J
f_{JB}^{I}
L^B_s. \nn
\eea
These equations can easily be
integrated since
\be
\partial^2_s\pmatrix{L_s \cr L_s^*}^{\bar I}=({\cal M}^2)^{\bar I}_{\bar
J}
\pmatrix{L_s \cr L_s^*}^{\bar J}
\ee
with
\be
({\cal M}^2)^{\bar I}_{\bar J}=\frac{1}{4}
\pmatrix{f_{B\bar K}^{\bar I}\theta^{\bar K}\bar \theta^{L}f_{L\bar J}^B
&
  -  f_{B\bar K}^{\bar I}\theta^{\bar K}\bar \theta^{*L}f_{L\bar J}^{*
B}
    \cr
-f_{B\bar K}^{*  \bar I}\theta^{*  \bar K}\bar \theta^{  L}f_{L\bar J}^{
B} &
    f_{B\bar K}^{*  \bar I}\theta^{* \bar K}\bar \theta^{*L}f_{L\bar
J}^{*  B}
}
\ee
The solution to (\ref{diffs}) is then given by
\bea
\pmatrix{L \cr L^*}_s^{\bar I}& = &\left(\frac{\sinh s{\cal M}}{{\cal
M}}\right)_{\bar J}^{\bar I}
\pmatrix{D\theta \cr D\theta^*}^{\bar J} \nn \\
L^A &=& e^A-\frac{1}{2}\pmatrix{\bar \theta^{I}f_{I\bar J}^{A}, &
 -\bar \theta^{*I}f_{I\bar J}^{*A}} \left(\frac{\sinh^2 (s{\cal
M}/2)}{{(\cal
 M})^2}\right)^{\bar I}_{\bar K}\pmatrix{D\theta \cr D\theta^*}^{\bar K}
 \label{vielbein}
\eea
with
\be
D^{IJ}=\d^{IJ}(d+\frac{1}{4}\omega^{ab}\g_{ab}
+\frac{1}{4}\omega^{a'b'}\g_{a'b'}) +\e^{IJ}\frac{i}{2}
(e^a\g_a -i e^{a'} \g_{a'})  \label{Dtheta}.
\ee
Here, we used the initial conditions
\be
L^a(\theta=0)=e^a, \ L^{a'}(\theta=0)=e^{a'},\ L^{ab}(\theta=0)=\omega^{ab},
\ L^{a'b'}(\theta=0)=\omega^{a'b'}.
\ee
The real vielbeins are then obtained by setting $s=1$.

Another virtue of above procedure is that one can obtain
the Wess-Zumino term as a world-sheet integral of an
expression which is itself integrated over $s$ \cite{D3}.
The important point is that
\be
\partial_{s}{\cal H}_{3s}=d\Omega_{2s}
\ee
where ${\cal H}_s$ is obtained from ${\cal H}$ by
replacing all $L$ by $L_s$, and where
\be
\Omega_{2s}=\frac{i}{2}s^{IJ}\left(\bar \theta^I \g_a\wedge L_s^J\wedge
L_s^a+i\bar \theta^I
\g_{a'} \wedge L_s^J\wedge L_s^{a'}\right)+c.c.
\ee
This can be verified with the differential equations (\ref{Maurer})
and (\ref{diffs}).
Hence
\be
S_{WZ}=\int_{M^3} {\cal H}_{3s}|_{s=1}=\int_{M^2}\int_{s=0}^1
\Omega_{2s}. \label{simpleWZ}
\ee

\section{Simplification of the Action}
\setcounter{equation}{0}

We now turn to the very important aspect of simplifying
the action. We will follow here the ideas of \cite{SuperKilling,ads5}
and fix $\kappa$-symmetry in the background adapted way. The procedure
consists of two steps:
\begin{itemize}
\item Choosing the  gauge
\be
\theta_-^I\equiv {\cal P}_-^{IJ} \theta^J\equiv
\frac{1}{2}\left(\d^{IJ}-i\e^{IJ}\G^0\G^1\right)\theta^J=0
\label{Projection}
\\
\ee
and
\item  redefining  the remaining fermions $\theta_+$ to
be space-time dependent as
\be \theta_+^I(x)=g_{tt}^{1/4}\vt_+^I \label{redefine}
\ee
where $\vt_+^I$ are  constant spinors which satisfies also ${\cal
P}_-^{IJ}\vt_+^J=0$.
\end{itemize}
This gauge is motivated by the observation
 that $D\theta$ as defined in (\ref{Dtheta}) is essentially
simply the Killing equation on $AdS_3\times S^3$ augmented
by a fermionic differential operator $d\vt \partial_\vt +
\bar d\vt \partial_{\bar \vt}$.
Hence, choosing the fermionic coordinates $\theta$ in (\ref{prop})
to be
space-time dependent Killing spinors, i.e.
\be
 \theta^{I\a\a'}(x)=e^{I\a\a'}_{\underline{J\b\b'}}(x)\vt^{\underline{J\b\b'}}
 \label{redef}
\ee
where $e^{I\a\a'}_{\underline{J\b\b'}}$ is a known space-time dependent
matrix and $\vt=$const, leads to
\be
D\theta^I=e^{I}_{\underline{J}}(x)d\vt^{\underline{J}}.
\ee
The Killing spinors on $AdS_3\times S^3$ in horospherical coordinates
can, for example, be found in \cite{KillingSpinors}, and it can
be easily verified, as first noted in \cite{FixKilling}, that
using $\kappa$-symmetry to project on half of them
precisely via (\ref{Projection}) leads to the fact that $\theta_+$ and
$D\theta_+$ obey the same projection, i.e.
\be
{\cal P}_-\theta_+={\cal P}_-D\theta_+=0,
\ee
since (\ref{redef}) reduces for this component to\footnote{Incidentally,

the surviving $\theta_+(x)$
spinor is nothing but the Killing spinor of the full D1-D5 geometry,
in the near horizon region. This might have some so far not understood
implications.}
\be
 \theta_+^I(x)=g_{tt}^{1/4}\vt_+^{J}.
\ee
Since this gauge is based on the isometry of the background, it
is  called the killing spinor gauge and was proposed
in \cite{FixKilling} as a procedure to gauge-fix $\kappa$-symmetry
of extended objects in their own background. In \cite{ads5} it
was shown that it could also be used to simplify dramatically the GS
string action on $AdS_5\times S^5$. Since the arguments given there
for admissibility of the gauge are exactly the same needed here
we refer the reader to that publication.

What we will show now is that with this gauge we have
\be
{\cal M}_+^2 \pmatrix{ D\theta_+ \cr D\theta_+^*}=0
\ee
which clearly simplifies (\ref{vielbein}) and
therefore the action dramatically. The important
fact to use is that terms of the form
\be
\bar \theta_+^I\hat\G D\theta_+^I, \qquad
{\rm with }\qquad [\G^{01},\hat \Gamma]=0 \label{vanish}
\ee
vanish. This implies that
\be
f_{B\bar K}^{\bar I}\theta^{\bar K}_+ \bar \theta^{L}_+ f_{L\bar K}^B
d\theta^{\bar K}_+=
f_{i\bar K}^{\bar I}\theta_+^{\bar K} \bar \theta_+^{L} f_{L\bar K}^i
D\theta^{\bar K}_++f_{(i2)\bar K}^{\bar I}\theta^{\bar K}_+ \bar
\theta^{L}_+ f_{L\bar K}^{(i2)}
D\theta^{\bar K}_+
\ee
with $i=0,1$, i.e., in the sum over the bosonic generators $B$
only the two momenta $P_i$ and the two Lorentz generators
$J_{i2}$ can contribute. Then, with a little algebra and using the
explicit form of the structure constants we find that in fact
the contributions from $P_i$ and $J_{i2}$ arise with opposite
sign and cancel. The same happens for the other term, i.e.
\be
f_{B\bar K}^{\bar I}\theta_+^{\bar K}\bar \theta_+^{*L}f_{L\bar J}^{*
B}
D\theta^{*\bar J}_+=0.
\ee
Putting all this  we see indeed that

\vspace{3mm}

\be
({\cal M}_+^2)^{\bar I}_{\bar J}\pmatrix{D\theta_+^{\bar J}\cr
D\theta_+^{*\bar J}}
=\pmatrix{
f_{B\bar K}^{\bar I}\theta_+^{\bar K}\bar \theta_+^{L}f_{L\bar J}^B
D\theta_+^{\bar J}
-f_{B\bar K}^{\bar I}\theta_+^{\bar K}\bar \theta_+^{*L}f_{L\bar J}^{*
B} D\theta_+^{*\bar
J}\cr
-f_{B\bar K}^{*\bar I}\theta_+^{*\bar K}\bar \theta_+^{*L}f_{L\bar
J}^{*B}  D\theta_+^{*\bar J}
+f_{B\bar K}^{*\bar I}\theta_+^{*\bar K}\bar \theta_+^{L}f_{L\bar J}^{B}
D\theta_+^{\bar
J}}=0.
\ee

\vspace{5mm}

\noindent
Now, recall that one
explicit form of the $AdS_3\times S^3$
metric in the "$2+4$"-split is
\be
ds^2=y^2(dx^p dx_p)+\frac{1}{y^2}(dy^t dy^t)
\ee
where $t$ and $p$ denote coordinate transverse $(y^2,y^3,y^4,y^5)$
and parallel $(x^0,x^1)$ to the brane.
With this form of the metric,   (\ref{redefine}) and
\be
\vt\equiv\vt^1
\ee
we find the simple supervielbeins
the supergeometry reads
\bea
(L_{s}^{ I})_+ &=&s \sqrt{|y|} d\vt^I \nonumber \\
(L_{s}^{I})_- &=&0
\nonumber \\
L_s^{p}&=&|y|(dx^{\hat m} -  \frac{is^2}{2} (\bar
\vt\gamma^{p} d\vt-d\bar
\vt\gamma^{p} \vt))   \label{La2} \\
L_s^{t}&=&\frac{1}{|y|} dy^{t}. \nonumber
\end{eqnarray}
Finally, inserting this into (\ref{action}) we obtain
\begin{eqnarray}
S =-\frac{1}{2}\int d^2\sigma\ \biggl[\sqrt{g} \, g^{ij}&&
\hspace{-0.7cm}
\biggr(y^2(\partial_i x^p
-  \frac{i}{2} (\bar
\vt \Gamma^{p} \partial_i\vt - \partial_i\bar
\vt \Gamma^{p} \vt))\times \nn \\
& & \hspace*{-3mm} \times (\partial_j x_p - \frac{i}{2} (\bar
\vt \Gamma_{p} \partial_j \vt-\partial_j\bar
\vt \Gamma_{p}  \vt)) + \nn \\
&&+\frac{1}{y^2} \partial_i y^t
\partial_j y^t \biggr) +\nonumber \\ &&\hspace*{-16mm}
 -  \frac{1}{2}\e^{ij} \partial_i y^t (\bar \vt \Gamma^t \partial_j\vt
 -
 \partial_j\bar \vt \Gamma^t\vt
 )\biggr]
\label{SimpleAction}
\end{eqnarray}

\section{Conclusions and Open Questions}

We presented the action of the the string in an $AdS_3\times S^3$
background. We explicitly constructed the Wess-Zumino term as a
closed three-form from first principles by employing the
supercoset structure of the background geometry. It was then shown
that the action can be simplified significantly to contain
fermionic terms only up to quadratic order. Of course, it
is still non-linear and a quantization procedure is not
apparent off-hand.

Since the pure NS background can be solved explicitly in
the RNS formalism \cite{Giveon},
at least in that case one should be able to quantize the GS action
as well. The quantization procedure is not, however, obvious.
An approach to the problem may be to construct the currents corresponding
to the spacetime Virasoro algebra and comparing these to
those obtained from the RNS formalism.

Furthermore, from knowing the NS background, several things about
the RR background can be deduced, e.g. the spectrum of chiral
primaries. It is of great interest to see if these can be
computed directly from the string action.
\bigskip

\bigskip

\noindent
{\bf  Acknowledgements:}

We had stimulating discussions with Renata Kallosh.
J.R. is supported by NSF grant PHY-9219345 and A.R.
is supported by DOE grant DE-FG02-96ER40559.
\bigskip

\bigskip

\noindent
{\bf  Note Added:}

After completion of this work we became aware of
the paper by I. Pesando \cite{Pesando}
which has some overlap with the
present publication.

\setcounter{section}{0}
\setcounter{subsection}{0}

\appendix{ }

We start with the $SU(1,1|2)$ algebra in the form of
\cite{Claus}:
\bea && \left[D,P\right]=P, \qquad
\left[D,K\right]=-K, \qquad \left[K,P\right]=2 D \nn \\
&& \left[N_{mn},N_{pq}\right]= \d_{np}N_{mq}+\d_{mq}N_{np}-
 \d_{nq}N_{mp}- \d_{mp}N_{nq} \nn \\
&&\left[D,Q_i\right]=\frac{1}{2}Q_i,\qquad
\left[D,S_i\right]=-\frac{1}{2}S_i
\nn \\
&& \left[N_{mn},Q_i\right]=-\frac{1}{4}\g_{mn}Q_i, \qquad
 \left[N_{mn},S_i\right]=-\frac{1}{4}\g_{mn}S_i, \label{su112} \\
& & \{Q_{i\a'},Q_{j\b'}\}=-2\e_{ij}C_{\a'\b'} P, \quad
\{S_{i\a'},S_{j\b'}\}=-2\e_{ij}C_{\a'\b'} K \nn \\
&&\{Q_{i\a'},S_{j\b'}\}=-2\e_{ij}C_{\a'\b'}
D+\e_{ij}(\g^{mn})_{\a'\b'}N_{mn} \nn
\eea
Here, we have $i,j=1,2, \ \a',\b'=1,2$ are $SO(3)$ spinor
indices and $m,n=1,2,3$.
The $AdS_3\times S^3$ geometry is the supercoset manifold
$\frac{SU(1,1|2)^2}{SO(1,2)\times SO(3)}$ with bosonic
subgroup $\frac{SO(2,2)\times SO(4)}{SO(1,2)\times SO(3)}
\sim \frac{SO(1,2)^2\times SO(3)^2}{SO(1,2)\times SO(3)}$.
The strategy is to combine two copies of above algebra
(variables $X$ and $\tilde X$) and combine the spinors $Q,\tilde Q,
S,\tilde S$
into suitable $SO(2,2)\times SO(4)$-spinors and the
bosonic operators as generators of this group. We will then
convert the bosonic and fermionic generators to covariant
$6D$ objects which results in (\ref{algebra}).

We start with the bosonic $SO(1,2)$ subalgebra:
\bea
[D,P]=P\quad [K,P]=2D\quad [D,K]=-K.
\eea
which can be rewritten with $P_+={1\over 2}(P+K), P_-={1\over 2}(P-K)$
as
\bea
[D,P_+]=P_-\quad [D,P_-]=P_+\quad [P_+,P_-]=D.
\eea
These generators should be combined with their
counterparts $\tilde D, \tilde P$ and $\tilde K$ satisfying
the same algebra into one $SO(2,2)$ matrix $M_{AB}$.
One finds that with
\bea
M_{12}=i(D-\tD)   \quad M_{23}= P_-+\tP_- \quad M_{13}=-i(P_+-\tP_+)
\nn \\
M_{03}=i(D+\tD) \quad M_{01}=P_--\tP_-\quad M_{02}=-i(P_++\tP_+)
\eea
$M_{AB}$ satisfies indeed the proper SO(2,2) algebra:
\bea
[M_{ab},M_{cd}]=\eta_{ac}M_{db}-\eta_{ad}M_{cb}-\eta_{bc}M_{da}+\eta_{bd}M_{ca}
\eea
with the signature (-++-) for indices (0123).

We now turn to unifying the spinors $Q,\tilde Q, S,\tilde S$.
It is useful to keep the index structure of the $\g$ matrices and
spinors in mind:
\be
\g_{\a}^{\ \ \b}, \qquad {Q_\a}, \qquad {\hat Q^\a},
\ee
where $\hat Q$ is the $SO(2,2)$ conjugate spinor of
$Q$, i.e. $\hat Q\equiv Q^\dagger \G^0\G^3$.
With the following set of definitions (and the
convention that we take $SO(2,2)$ spinors to be
Majorana):
$$
\G^0=
\pmatrix{
0 & 1 \cr
-1 & 0 \cr
}\quad
\G^{1,2}=
\pmatrix{
0 & \s^{1,2} \cr
\s^{1,2} & 0 \cr
} \nn
$$
\be
\G^3=
\pmatrix{
0 & i\s^3 \cr
i\s^3 & 0 \cr
}, \quad \G^5=\pmatrix{
1 & 0 \cr
0 & -1 }, \quad
C=\G^0\G^2
\ee
we find
\be
M_{AB}\G^{AB}C=2\pmatrix{-2i K & 2i D & 0 & 0 \cr
                          2iD & -2iP & 0 & 0 \cr
                          0 & 0 & -2i\tP & 2i\tD \cr
                          0 & 0 &  2i  \tD & -2i\tK}
\ee
which reveals that one part of the algebra is given by
\be
 \{q_i,q_j\}=-\frac{i}{2}\e_{ij}M_{AB}\G^{AB}CC'+...
\ee
with
\be
q_{i\a'}=\pmatrix { S_{\a'} \cr -Q_{\a'}\cr -\tQ_{\a'}\cr
\tS_{\a'}}_i.
\ee
To complete the algebra we turn to the $SO(4)$ part, where spinors
are taken to be symplectic-Majorana. Our conventions are
\be
\G'^0=
\pmatrix{
0 & -i \cr
i & 0 \cr
}\quad
\G'^{a}=
\pmatrix{
0 & \s^{a} \cr
\s^{a} & 0 \cr
}\quad
 C'=\G'^0\G'^2, \quad \G'^5=\pmatrix{
1 & 0 \cr
0 & -1 }
\ee
and we find
\be
M'_{A'B'}\G'^{A'B'}C'=\pmatrix{-2 M'_{0'i'}\s^{i'} +M'_{i'j'}\s^{i'j'}
&  0 \cr
                                 0 & 2 M'_{0'i'}\s^{i'}
+M'_{i'j'}\s^{i'j'}}
C'
\ee
which implies that the $N_{i'j'}$ in (\ref{su112}) are given
by
\be
N_{i'j'}=\frac{1}{2}\left(M'_{i'j'}-\e_{i'j'k'}M'_{0'k'}\right)
\ee
With these preliminaries the $SO(2,2)\times SO(4)$
spinors are defined as
\be
q_{iI\a I'\a'}=\pmatrix{S_{1'\a'}\cr
                        S_{2'\a'}\cr
                        -Q_{1'\a'}\cr
                        -Q_{2'\a'}\cr
                        -\tQ_{1'\a'}\cr
                        -\tQ_{2'\a'}\cr
                        \tS_{1'\a'}\cr
                        \tS_{2'\a'}}_i
\ee
where the vector components denote the $q_{I\a}$ elements in the
natural order. The pair $I,\a$ ($I',\a'$) is an $SO(2,2) (SO(4))$
index, whereas
$i$ is still the symplectic index. Counting the
degrees of freedom reveals that half of the 32 components of
$q$
("real" by Majorana/symplectic-Majorana condition) have to be
projected out. The underlying reason
is that spinors transform under $SO(2,2)\times SO(4)\sim
SO(1,2)_1\times SO(1,2)_2\times SO(3)_1\times SO(3)_2$ only
under $SO(1,2)_1\times SO(3)_1$ or  $SO(1,2)_2\times SO(3)_2$,
since the algebra is the product $SU(1,1|2)^2$. Clearly, the
projector ${\cal P}$ has to ensure that $I=I'$ which results in
\be
{\cal P} =\frac{1}{2}(1\otimes 1' +\G^5\otimes \G'^5)
\ee
With these conventions the algebra reads
\bea
\{q_i,q_j\}&=&-\frac{i}{2}\e_{ij}{\cal P}
        \left(M_{AB}\G^{AB}C\otimes C'-C\otimes M'_{A'B'}\G'^{A'B'}C'\right) \nn
        \\
\left[M_{AB},q_i\right]&=&-\frac{1}{2}\G_{AB} q_i \\
\left[M'_{A'B'},q_i\right]&=&-\frac{1}{2}\G'_{A'B'} q_i \nn
\eea
plus the conventional $SO(2,2)$ and $SO(4)$ pieces.

In order to achieve more closeness to $6D$ quantities
it is useful to define
\be
\hat q_i\equiv \e_{ij}q_j^T C \otimes C',
\ee
which are the conjugate spinors since by
the symplectic-Majorana condition
we have
\be
(q_i)^*=\e_{ij}B\otimes B' q_j,
\ee
and to consider as fundamental supercharges
$q\equiv q_{i=1}$ and $\hat q\equiv \hat q_{i=1}$.
It is also convenient for
later purposes to change the basis  to
\be
\G^0\rightarrow\pmatrix {i & 0 \cr 0 & -i},
\qquad
\G'^0\rightarrow\pmatrix {1 & 0 \cr 0 & -1},
\ee
The components of $q$ in this basis  which
survive the projection are
\be
Q_1\equiv q_{11'}+q_{22'} \qquad
Q_2\equiv q_{12'}-q_{21'},
\ee
where the indices denote $I,I'$.
Although this is a source for confusion, let us denote
these generators by $Q_I$, where $I$ is not to be
confused with the $SO(2,2)$ index.
It is of crucial importance for symmetry considerations
to know that
\be
\hat Q_I=(\hat Q)_I=\hat q_{1I'}+\e_{I'J'}\hat q_{2J'}=+\e_{IJ} Q_J^\dagger \s_3 \label{symbar}
\ee
where $\s_3$ acts on the $\a$ index of $Q^\dagger$.
With $P_a,P_{a'}, J_{ab}, J_{a'b'}$ as defined in
(\ref{PJM})
we can  write down the
algebra
\bea \{Q_I, \hat
Q_J\}&=-i&\hspace{-3mm}\delta_{IJ}\left(J_{ab} \gamma^{ab} - J_{a'b'}
\gamma^{a'b'}\right)+ \nn \\ & & \qquad \qquad +2i\e_{IJ}\left( i
P_{a}\gamma^a - P_{a'}\gamma^{a'}\right)
\eea
\bea
\left[P_{a},Q_I\right]=-\frac{i}{2} \e_{IJ}\gamma_a Q_J  & &
\left[P_{a'},Q_I\right]=\frac{1}{2} \e_{IJ}\gamma_{a'} Q_J \nn \\
\left[M_{ab},Q_I\right]=-\frac{1}{2} \gamma_{ab} Q_I &&
\left[M_{a'b'},Q_I\right]=-\frac{1}{2} \gamma_{a'b'} Q_I \nn \\
\left[P_{a},\hat Q_I\right]=\frac{i}{2} \hat Q_J
\e_{JI}\gamma_a  &&
\left[P_{a'},\hat Q_I\right]=-\frac{1}{2} \hat Q_J
\e_{JI} \gamma_{a'} \nn \\
 \left[M_{ab},\hat Q_I\right]=\frac{1}{2}
\hat Q_I\gamma_{ab} && \left[M_{a'b'},\hat Q_I\right]=\frac{1}{2}
\hat Q_I \gamma_{a'b'} \label{gralgebra}
\eea
\bea
\left[M_{AB},M_{CD}\right]&=&\eta_{BC} M_{AD}+\eta_{AD}
M_{BC}-\eta_{AC} M_{BD}-\eta_{BD} M_{AC}\nn \\
\left[M_{A'B'},M_{C'D'}\right]&=&\d_{B'C'} M_{A'D'}+\d_{A'D'}
M_{B'C'}- \nn \\ & & \hspace*{2cm} -\d_{A'C'} M_{B'D'}-\d_{B'D'}
M_{AC} \nn \eea
In verifying the Jacobi identities, heavy use was made
of the following identities:
\bea (\s^a)_\a^{\ \g} (\s^a)_\b^{\
\d} & = & 2\d_\a^{\ \d} \d_\b^{\ \g} -\d_\a^{\ \g}\d_\b^{\ \d} \\
(\g^{ab})_\a^{\ \g} (\g_{ab})_\b^{\ \d} & = & -4\d_\a^{\ \d}
\d_\b^{\ \g}+2\d_\a^{\ \g}\d_\b^{\ \d} \label{Fierz} \eea
So far, the $6D$ covariance of the algebra is not quite
obvious. However, if define the $6D$ gamma matrices as in
(\ref{gammamat}), the chiral $6D$ supercharges $Q$  as in
(\ref{6dcharge}) and $\bar Q$ as in (\ref{barQ}) we find
from (\ref{gralgebra}) precisely (\ref{algebra}).
To see this it is noteworthy that $\hat Q$ and $\bar Q$
are related by
\be
\hat Q_I=-i\e_{IJ}\bar Q_J.
\ee


\end{document}